\documentclass[aps,prl,twocolumn,secnumarabic,amssymb, amsmath,nobibnotes,showpacs]{revtex4-1}
\usepackage{amsmath,hyperref}

\usepackage{bm}%
\usepackage{graphicx,calc,epsfig,bbm}
\usepackage{tikz}
\expandafter\ifx\csname package@font\endcsname\relax\else
 \expandafter\expandafter
 \expandafter\usepackage
 \expandafter\expandafter
 \expandafter{\csname package@font\endcsname}%
\fi

\newcommand {\be}{\begin{equation}}
\newcommand {\ee}{\end{equation}}
\newcommand{\ba}{\begin{array}{c}}
\newcommand{\ea}{\end{array}}

\begin{document}
\title{Quantum-Reduced Loop-Gravity: Relation with the Full Theory}%

\author{Emanuele Alesci$^{1}$, Francesco Cianfrani$^{2}$ and Carlo Rovelli$^{3,4}$}%
\affiliation{$^{1}$Instytut Fizyki Teoretycznej, Uniwersytet Warszawski,ul. Ho$\dot{z}$a 69, 00-681 Warszawa, Poland, EU\\
$^{2}$Institute
for Theoretical Physics, University of Wroc\l{}aw, Pl.\ Maksa Borna
9, Pl--50-204 Wroc\l{}aw, Poland.\\
$^{3}$ Aix Marseille Universit\'e, CNRS, CPT, UMR 7332, 13288 Marseille, France.\\
$^{4}$ Universit\'e de Toulon, CNRS, CPT, UMR 7332, 83957 La Garde, France.
}
\date{\today}%

\begin{abstract} 
\noindent The quantum-reduced loop-gravity technique has been introduced for dealing with    cosmological models. We show that it can be applied rather generically: anytime the spatial metric can be gauge-fixed to a diagonal form. The technique  selects states based on reduced graphs with Livine-Speziale coherent intertwiners and could simplify the analysis of the dynamics in the full theory.     
\end{abstract}


\maketitle

Quantum Reduced Loop Gravity (QRLG) is a framework introduced for the quantization of symmetry-reduced sectors of general relativity.  It was introduced in \cite{Alesci:2012md,Alesci:2013xd} and applied to an inhomogeneous extension of the Bianchi I cosmological model.  Here we show that its application is in fact quite wide, since it essentially amounts to a choice of gauge in the full theory. More precisely, we show that fixing the gauge where the triad is diagonal (in the quantum theory) leads to the  state space of QRLG. 

The loop quantization of homogeneous models \cite{Bojowald:2011zzb,Ashtekar:2011ni} (loop quantum cosmology) and spherically-symmetric systems \cite{Gambini:2013qf} (black holes) has been mostly studied by first restricting to a reduced phase space and then quantizing the resulting system.  The strategy of starting from the full quantum theory and restricting the set of states has  developed more slowly, both in the canonical \cite{Rovelli:2008dx} and covariant \cite{Bianchi:2010zs,Vidotto:2010kw,Borja:2011ha} versions of the theory.

The metric of a Bianchi I model is diagonal and the internal $SU(2)$ gauge can be fixed so that the densitized triads are diagonal as well. In QRLG, one fixes a three-dimensional cubic lattice oriented in the directions that diagonalize the metric.  The connection on each link belongs then to a fixed $U(1)$ subgroup of $SU(2)$, one per each of the three possible orientations of the links. Group elements associated to links are in $U(1)$, not in $SU(2)$, and the $SU(2)$ structure is only present at the nodes. The way $U(1)$ states are embedded into $SU(2)$ states is analogous to the way  $SU(2)$ states sits into $SL(2,C)$ states in spinfoam theory,  one dimension up.  Using these structures we can regularize the scalar constraint as in full  loop quantum gravity  (LQG)\cite{Thiemann96b}.   

Here we point out that this scheme is  more general than its application to Bianchi I and the inhomogeneous extensions previously considered. It  works anytime we can choose a reference frame where the spatial metric is diagonal.  This is generically possible, since any 3-metric can generically be taken to diagonal form by a 3d diffeomorphism  \cite{Grant:2009zz}, as in three dimensions the number of nondiagonal components of the metric coincides with the number of parameters of a diffeomorphisms. The price to pay is a nontrivial Shift function and  a potentially more complicated dynamics.  

Below, we review a few basis elements of LQG that we need for this construction, then we give the QRLG construction of the state space, and finally we recover this same  state space by a gauge fixing in the general quantum theory.

\paragraph{Loop Quantum Gravity.} In LQG, the elements of the kinematical Hilbert space $\mathcal{H}^{kin}$ are labeled by oriented graphs $\Gamma$ in the spatial manifold and are given by functions on $L$-copies of $SU(2)$, $L$ being the number of links in $\Gamma$. A basis of states is obtained from Peter-Weyl theorem, and is labelled by an irreducible representations ${\bf j_l}$ of $SU(2)$ on each link $l$, and a $SU(2)$ intertwiner $x_n$ at each node $n$. The corresponding state reads  
 \be
<h_l |\Gamma,{\bf j_l},{\bf x_n}>=\prod_{n\in\Gamma} x_{n}\cdot   \prod_{l\in\Gamma}D^{j_{l}}(h_{l}).
\label{spinnet solite} 
\ee 
$D^{j}(h)$ and $x$ are Wigner matrices in the representation $j$ and intertwiners, respectively; the products extend over all the links and the nodes in $\Gamma$; the dot means the contraction between  indices of  intertwiners and  Wigner matrices. The flux operator $E_i(S)$ associated to the oriented surface $S$ acts as the left (right) invariant vector fields on the group element based at  links $l$ beginning (ending) on $S$.  For instance, given a surface $S$ having a single intersection with a link $l$ at a point $x\in e$, such that $l=l'\bigcup l''$ and $l'\cap l''=x$, the operator $\hat{E}_i(S)$ is given by 
\be
\hat{E}_i(S)D^{j_l}(h_l)
=8\pi\gamma l_P^2 \; o(l,S) \; D^{j_l}(h_{l'})\,\tau_{i}\,D^{j_l}(h_{l''}),\label{Eop}
\ee  
$\gamma$ and $l_P$ being the Immirzi parameter and the Planck length, respectively, while $o(l,S)$ is equal to $0,1,-1$ according to the relative sign of $l$ and the normal to $S$. $\tau^i$ denotes the $SU(2)$ generators in the $j_l$ representation.

The equivalence class $s$ of graphs $\Gamma$ under diffeomorphisms can be used to implement background independence in the dual of the $SU(2)$-invariant kinematical Hilbert space as follows
\be\label{sk}
<h |s,{\bf j_l},{\bf x_n}>=\sum_{\Gamma\in s}<h |\Gamma,{\bf j_l},{\bf x_n}>^*.
\ee 

The scalar constraint can be regularized in the space of SU(2)- and diffeo-invariant states.

\paragraph{Quantum Reduced Loop Gravity.} The Bianchi I model is endowed with a diagonal metric tensor
\be \label{dm}
dl^2=a_1^2(dx^1)^2+a_2^2(dx^2)^2+a_3^2(dx^3)^2,
\ee
where $a_i$ ($i=1,2,3$) are three time-dependent scale factors. In the inhomogeneous extension of Bianchi I, the $a_i$ are assumed to be a function of  time and the spatial coordinates $x^i$, which are the Cartesian coordinates of a fiducial flat metric. The associated densitized triads can be chosen to be diagonal, {\it i.e.}
\be
E^a_i=p^i\delta^a_i,\quad |p^i|=\frac{a_1a_2a_3}{a_i}
\ee
by the gauge-fixing condition \cite{Cianfrani:2011wg,Cianfrani:2012gv}
\be\label{chi}
\chi_i=\epsilon_{ij}^{\phantom{12}k}E^a_k\delta^j_a=0.
\ee
The connections are generically given by 
\be\label{conn}
A^i_a=c_i u^i_a+\ldots,\quad c_i=\frac{\gamma}{N}\dot{a}_i,
\ee
where $u_a^i = \delta_a^i$ are the component of three units vectors $\vec u_a$ oriented along thee fiducial orthogonal axis and the dots indicate terms due to the spin connections, which are generically non-diagonal. These terms were  disregarded in \cite{Alesci:2012md,Alesci:2013xd} by considering two  cases: the reparametrized Bianchi I model, in which each $a_i$  is a function of the single corresponding coordinate $x^i$; and the Kasner epoch inside a generic cosmological solution, for which spatial gradient of the scale factors are negligible with respect to time derivatives. 

The kinematical symmetries in this reduced phase space are generated by two sets of constraints: the Gauss constraints associated with three $U(1)$ gauges, each acting on a single spatial direction $x^i$ and having $\{c_i,p^i\}$ as the couple of connections and momenta; the vector constraints associated with a subgroup of the diffeomorphisms group, made by those transformations (reduced diffeomorphisms) which can be seen as the product of a generic diffeomorphisms along a given direction $x^i$ and a rigid translation along the other ones.  

The description of such a system in QRLG is obtained by truncating the LQG kinematical Hilbert space. First, the Hilbert space of the full theory is restricted to that based on a reduced set of cubic graphs, with links parallel to three fiducial vectors $\omega_i=\delta_i^a\partial_a$ ($i=1,2,3$).  We call $i_l$ the direction of the link $l$ in the cubic graph.  

Then, the gauge fixing leading to diagonal momenta and connections is implemented weakly, following the procedure to impose the simplicity constraints in Spin Foam \cite{Engle:2007wy}.  The condition (\ref{chi}) is first rewritten in terms of fluxes across surfaces $S^j$normal to the $j$ direction, as 
\be
\chi_i(S)=\epsilon_{ij}^{\phantom{12}k}E_k(S^j)=0
\ee
and then implemented solving strongly the master constraint condition $\hat\chi^2(S)=\sum_{i}\hat\chi_i^2(S)=0$. Since the holonomy along the link $l$, is generated by $\tau_{i_l }$ only, 
\be
h_{l}=Pe^{ \left( \int_{l} c_i dx^i\right)\tau_{i_l}},
\ee
and a solution of $\hat{\chi}^2(S)\tilde\psi_{l}=0$ can be obtained by working with projected $U(1)$-states, obtained by stabilizing the $SU(2)$ group element based at each link $l$ around the internal directions $\vec{u}_l$, where $\vec{u}_{l}=\vec{u}_{i_l}$ and the components of $\vec{u}_i$ are given by  $u_a^i = \delta_a^i$ as above.  In terms of Wigner matrices, the resulting projected state on a link with direction $i=1,2,3$ reads
\be
\tilde{\psi}_i(h)= \sum_{n=-\infty}^{+\infty} \psi^{n}\  {}^i\!D^{j(n)}_{n\,n
}(h), 
\label{projected 4}
\ee
where ${}^i\!D^{j(n_i)}_{mr}$ are the Wigner matrices in the spin basis $|j,m\rangle_i$ that diagonalizes the operators $J^2$ and $J_i$, and $\psi^n$ are the coefficients of the expansion. The condition $\hat{\chi}^2\tilde\psi_{e_i}=0$ fixes the degree of the representation, {\it i.e.} the $U(1)$ quantum number $n$ in terms of  $SU(2)$ quantum number $j$. An approximate solution which becomes exact for $j\rightarrow\infty$ is given by
\be \label{deg}
j(n)=|n|.
\ee
This is good enough for assuring the classical limit.  Here we restrict to positive values of $n$ for simplicity. Let's
$\mathcal{H}^{R}$ be the space spanned by the states (\ref{projected 4}), with $j$ given by
 \eqref{deg}. The gauge-fixing condition $\langle\hat\chi_i\rangle=0$ holds weakly on this space. 
 
A reduced recoupling theory adapted to such states follows from $SU(2)$ recouping theory. Consider the  $SU(2)$ coherent states 
\be
|j,\vec{u}\rangle=D^j(\vec{u})|j,j\rangle=\sum_m |j,m\rangle D^j_{mj}({u}), 
\ee
where $\vec{u}$ is a unit vector and $u$ is a group element that rotates the $z$ axis into $\vec{u}$. Using these, define the projectors  
\be\label{proj}
P_{l}= |j_l, \vec{u_l}\rangle \langle j_l, \vec{u_l}|, 
\ee
for each link of the graph. 

The projector $P_\chi$ that maps $\mathcal{H}^{kin}$ into $\mathcal{H}^{R}$ acts on each Wigner-matrix state as    
\be\label{proj1}
P_\chi: D^{j_l}(h_{l})\ \mapsto  \ P_l D^{j_l}(h_{l})P_l,
\ee
and its image has the form \eqref{projected 4}.

So far we have considered states on single links.  Now let's consider states of the full theory, invariant under SU(2) gauge transformations. The projection of the invariant basis states can be written in the form
\be
\langle h|\Gamma, {\bf j_l},{\bf x_n}\rangle_R= \prod_{n\in\Gamma} \langle {\bf j_{l}}, x_n|{\bf j_{l}},\vec{{\bf u}}_{\bf l} \rangle \,  \cdot \;  \prod_{l\in\Gamma}{}^{i_l}\!D^{j_l}_{j_lj_l}(h_{l}).
\label{base finale}
\ee
The coefficients $\langle{\bf j_l}, x_n|{\bf j_l}, \vec{{\bf u}}_{\bf l}\rangle$ are the reduced intertwiners and they take the following expression in terms of the $SU(2)$ intertwiner basis  
 \be
 \langle{\bf j_l}, x|{\bf j_l},\vec{{\bf u}}_{\bf l}  \rangle= x^{*}_{m_1... m_O, m'_1 ... m'_I }\prod^{O}_{o=1} D^{-1j_{o}}_{j_{o}m_o}(u_o) \prod^{I}_{i=1}  D^{j_{i}}_{m'_i j_{i}}(u_i),
\nonumber
 \ee
where we have split the links $l=\{i,o\}$ $i=1,...,I$, $o=1,...,O$ in $n$ into $I$ incoming and $O$ outgoing links. A generic state can thus be expanded as follows
\be\label{projst}
{}_{R}\langle\Gamma, {\bf j_l},{\bf x_n}| \psi\rangle = \prod_{n\in\Gamma} \langle {\bf j_{l}}, \vec{{\bf u}}_{\bf l} |{\bf j_{l}}, x_n \rangle\cdot \;  \prod_{l\in\Gamma} \; \psi^{j_{l}}_l\;.
\ee

The reduced intertwiners $\langle{\bf j_{l}}, x_n|{\bf j_{l}},\vec{{\bf u}}_{\bf l}  \rangle$ provide a non-trivial node structure. It is the presence of such a node structure and of reduced diffeomorphisms invariance which provide a well-defined regularized expression for the scalar constraint by mimicking the techniques of quantum spin dynamics \cite{Thiemann96b}.

We now re-interpret  restriction to reduced graphs as a gauge fixing at the quantum level, to a gauge where the metric tensor takes the form (\ref{dm}).  This turns out to be simpler than the Bianchi I case considered previously, because it does not require to chose a priori the projected form for the states; this form comes automatically from the gauge fixing. 

\paragraph{Fixing the frame.} 
Given a point $x$ and three vectors $\omega_i=\delta_i^a\partial_a$ at the point, let $S^i_x$ be three surfaces intersecting at $x$ dual to these vectors. The vanishing of the off-diagonal components of the metric tensor can be written in terms of fluxes as follows
\be\label{dgf}
\eta^{km}_x=\delta^{ij}E_i(S^k_x)E_j(S^m_x)=0,\qquad k\neq m,\quad \forall x\in\Sigma.
\ee  
Consider now the equation as a gauge-fixing constraint in the quantum theory. 
We want thus to solve  $\hat\eta^{kl}_x=0$, {\it i.e.} weakly. That is, we look for a subspace of the full Hilbert space where 
\be
\langle\psi|\eta^{km}_x|\phi \rangle=0,\qquad k\neq m,\quad\forall x\in\Sigma.
\ee     
There are two cases for which the action of the operator $\hat{\eta}^{kl}$ on a state based in $\Gamma$ is non-trivial. depending on the intersections between $\Gamma$ and the surfaces $S^i_x$:
\begin{enumerate}\addtolength{\itemsep}{-2mm}
\item there is an link $l_x\in \Gamma$ containing $x$ as an internal point; 
\item $x$ is a node for $\alpha$.
\end{enumerate} 
In the first case, the action of $\hat\eta^{km}_x$ is non-trivial on $D^{j_{l(x)}}(h_{l(x)})$ and reads
\begin{eqnarray}
 \hat\eta^{km}_xD^{j_{l_x}}(g_{l(x)})
&=&(8\pi\gamma l^2_P)^2\ \ \ \times 
\\ \nonumber 
&&\hspace{-2em} o(S^k,l_x)\ o(S^m,l_x) \ j_{l_x}(j_{l_x}+1)\ D^{j_{l_x}}(h_{l_x}),
\end{eqnarray}
where $o(S,l)$ is the intersection number between the link and the surfaces.  Hence,  spin networks with the link $l_x$ are eigenfunctions of the operator $\hat\eta^{kl}_x$. Therefore, the scalar product with other spin networks with a link $l_x$ gives 
\begin{eqnarray}
&&\langle l_x,\tilde{j}_{l_x}|\hat\eta^{km}_x|l_x,j_{l_x}\rangle=\\ \nonumber
&&\hspace{2em}
=(8\pi\gamma l^2_P)^2o(S^k,l_x)o(S^m,l_x)j_{l_x}(j_{l_x}+1)\delta_{\tilde{j}_{i},j_{i}},
\end{eqnarray}
which in general does not vanish for $\tilde{j}_{l_x}=j_{l_x}$. However, a proper subspace exists where all these matrix element vanish. It is formed by states based on the links of the cubic graph, {\it i.e.} at links parallel to the vectors $\omega_i$. In fact, if $l_x$ is in the direction $i=1,2,3$ then $o(S^k,l_x)=\delta^k_i$ and 
\be
\langle l_x,\tilde{j}_{l_x}|\hat\eta^{km}_x|l_x,j_{l_x}\rangle
=(8\pi\gamma l^2_P)^2 \delta^k_i \delta^m_i \, j_{l_x}(j_{l_x}+1)\delta_{\tilde{j}_{l_x},j_{l_x}}
\nonumber
\ee
which vanishes for $k\neq m$. Henceforth, the restriction to reduced graph satisfies (\ref{dgf}) in case 1.  We denote reduced graphs by $\Gamma_P$ and the Hilbert space based at $\Gamma_P$ by $\mathcal{H}_P$. 

We can then follow \cite{Alesci:2012md,Alesci:2013xd} and define a projector $P$  selecting the states based at reduced graphs and projecting to $\mathcal{H}_P$ diffeomorphisms invariant states (\ref{sk}). This gives 
\be
<h |P|s,{\bf j_l},{\bf x_v}>=
\sum_{\Gamma_P\in s}<h |\Gamma_P,{\bf j_l},{\bf x_n}>^*,
\ee    
where the sum is over all the reduced graphs contained in the $s$-knot $s$. Reduced graphs within each $s$ are mapped into each others by the action of reduced diffeomorphisms, times all  possible exchanges between fiducial vectors $\{\omega_i,-\omega_i\}$. Hence, $s$-knots are projected to sums of reduced $s$-knots $s^A_P$
\be
<h |P|s,{\bf j_l},{\bf x_n}>=\sum_A\sum_{\Gamma_P\in s^A_p}<h |\Gamma_P,{\bf j_l},{\bf x_n}>^*,
\ee   
with the index $A$ labeling all permutations of $\{\omega_i\}$ times inversions. The sum over $A$ implies us that no special meaning must be given to a fiducial direction.  

This solution to the gauge fixing condition (\ref{dgf}) defines the same Hilbert space as in QRLG with the only difference that we have to sum all permutations and inversions of the fiducial directions. 

Let us now move to case 2. Here a solution in the large $j$ limit is obtained restricting the admissible intertwiners states to the Livine-Speziale coherent intertwines \cite{Livine:2007vk} with normals $\vec u_l$. Livine-Speziale coherent states adapted to the reduced graphs are given by inserting a resolution of the identity
\be
\langle h| \Gamma,{\bf j_l},{\vec{\bf u}}_{\bf l}\rangle= \sum_{{\bf x_n}}\langle h|\Gamma, {\bf j_l},{\bf x_n}\rangle \langle  {\bf j_{l}},  {\bf x_n} | {\bf j_{l}}, \vec{{\bf u}}_{\bf l} \rangle.
\ee
The matrix elements of the product of two fluxes intersecting $\Gamma$ at a node $n$ for $j\rightarrow \infty$ are (see \cite{Bianchi:2009ri})
\begin{eqnarray}
&&\langle\Gamma,{\bf j_l},{\vec{\bf u}}_{\bf l}| \vec{E}(S^k_n)\cdot\vec{E}(S^m_n) |\Gamma,{\bf j_l},{\vec{\bf u}}_{\bf l}\rangle\approx
\\\nonumber &&\hspace{6em}
(8\pi\gamma l_P^2)^2 \sum_{l_k}j_{l_k}\vec{u}_{k}\cdot\sum_{l_m}j_{l_m}\vec{u}_{m}, 
\end{eqnarray}
where the sums extend over the links emanating from $n$ in the direction $\vec{u}_k$ and $\vec{u}_m$. Since the vectors $\vec{u}_i$ are orthogonal, the expression above vanishes for $k\neq m$. We have assumed for simplicity that all the links are outgoing. 
Therefore, the condition $\langle\eta_n^{km}\rangle=0$ can be solved in the large $j$ limit and it provides the restriction to the states of the form 
\be\label{ex}
\langle \Gamma, {\bf j_l},{\bf x_n}| \psi\rangle=\prod_{n\in\Gamma} \langle{\bf j_{l}}, {\vec{\bf u}}_{\bf l} | {\bf j_{l}}, x_n \rangle \prod_{l}\psi^{j_{l},\vec{u}_l}_{l},
\ee
in which $\psi^{j_{l},\vec{u}_l}_{l}$ denotes the coefficients of the expansion of the $SU(2)$ group elements in the basis of coherent states. By the identification 
\begin{eqnarray}
\psi^{j_{l},\vec{u}_l}_{l}&=&\psi^{n_{l}}_{l}, \quad \mathrm{for}\quad n_{l}= j_{l}, 
\end{eqnarray}
the expression (\ref{ex}) formally coincides with the one found in (\ref{projst}) giving the expansion of the states of QRLG in the basis elements of $\mathcal{H}^{R}$. However, now we have an actual expansion in the basis elements of $\mathcal{H}_P$, {\it i.e.} of the full theory just restricted to reduced graphs.

The $SU(2)$ gauge-fixing condition can also be imposed without using projected $U(1)$ networks.  As pointed out in  \cite{Alesci:2012md,Alesci:2013xd}, it is convenient to write Wigner matrices based at links in the direction $i$ in the basis $|j,m\rangle_i$ diagonalizing $J^2$ and $J^i$, so that the action of the master constraint condition $\hat\chi^2(S_x)=0$ at the node reads 
\be
\hat\chi^2(S_x){}^i\!D^j_{mn}(h_{l})=(8\pi\gamma l_P^2)(j(j+1)-m^2){}^i\!D^j_{mn}(h_{l}).
\ee
A solution for $j\rightarrow\infty$ is given by $m=j$ and can be implemented by inserting the projector $P_l$ at the node. 

The general reduced basis element is obtained from (\ref{spinnet solite}) replacing $D^{j_{l}}(h_{l})$ with $P_lD^{j_{l}}(h_{l})P_l$, and this gives  
\be
\langle h|\Gamma, {\bf j_l},{\bf x_n}\rangle=\prod_{n\in\Gamma} \langle {\bf j_{l}}, x_n|{\bf j_{l}},\vec{{\bf u}}_{\bf l} \rangle \prod_{l\in\Gamma}\;{}^l\!D^{j_{l}}_{n_{l} n_{l}}(h_{l}),
\ee
which coincides with Eq.(\ref{projected 4}). Hence, the quantum states adapted to the gauge fixing condition (\ref{chi}) coincide with the ones defined in \cite{Alesci:2012md,Alesci:2013xd} even if the connection is not diagonal.

\paragraph{Conclusions.} 

We have discussed how to fix a gauge where the triad is diagonal, in the kinematical Hilbert space of LQG.  We have shown that the gauge fixing condition is solved weakly by  states 
based at reduced links connected by Livine-Speziale coherent intertwiners. This leads to the same state space as the one defined in Quantum Reduced Loop Gravity (QRLG) of   \cite{Alesci:2012md,Alesci:2013xd}. 

Therefore, QRLG can be regarded as a framework useful beyond the cosmological context, possibly for full quantum gravity.  
The fact that the analytical expression for the hamiltonian constraint simplifies substantially in the QRLG language  \cite{Alesci:2012md} (essentially due to the fact that the volume is diagonal in the reduced basis elements) makes this result  intriguing.  The limits of the construction are in the approximated solution to the gauge fixing conditions, which holds only for $j\gg1$, possibly in the limitations of the applicability of the gauge condition, and perhaps in the complication of the dynamics that one might expect in a gauge fixed context like this. We expect the semiclassical analysis to indicate whether any interesting quantum gravity effects can be captured in this regime. The framework can also in principle simplify other issues, such as the coupling between quantum geometry and matter \cite{ThiemannBook,Bianchi:2010bn,Rovelli:2009ks} and the relation between the canonical and covariant approach \cite{Alesci:2011ia,Alesci:2013kpa}.

{\it Acknowledgment}   
FC is supported by funds provided by the National Science Center under the agreement
DEC-2011/02/A/ST2/00294. The work of E.A. was supported by the grant of Polish Narodowe Centrum Nauki nr DEC-2011/02/A/ST2/00300.
This work has been partially realized in the framework of the CGW collaboration (www.cgwcollaboration.it).


%

\end{document}